\documentclass[conference]{IEEEtran}
\usepackage[latin9]{inputenc}
\usepackage{amsmath}
\usepackage{amssymb}
\usepackage{graphicx}
\usepackage{fancyhdr}
 
\makeatletter
\IEEEoverridecommandlockouts
\usepackage{cite}
\usepackage{amsfonts}\usepackage{algorithmic}
\usepackage{subfigure}
\usepackage{textcomp}
\usepackage{xcolor}
\def\BibTeX{{\rm B\kern-.05em{\sc i\kern-.025em b}\kern-.08em
    T\kern-.1667em\lower.7ex\hbox{E}\kern-.125emX}}
\makeatother

\begin{document}

\title{Reduced Plasma Frequency Calculation Based on Particle-In-Cell Simulations }

\author{\IEEEauthorblockN{Tarek Mealy\textsuperscript{1}, Robert Marosi\textsuperscript{1},
and Filippo Capolino\textsuperscript{1,2}} \IEEEauthorblockA{\textit{(1) Department of Electrical Engineering and Computer Science,
University of California, Irvine, CA 92697, USA} \\
\textit{(2) Department of Communications and Signal Theory, University
Carlos III of Madrid, Spain} \\
%\textit{University of California, Irvine, CA 92697 USA}\\
tmealy@uci.edu, rmarosi@uci.edu, and f.capolino@uci.edu}}
\maketitle
\thispagestyle{fancy}
\begin{abstract}
\noindent We propose a scheme to calculate the reduced plasma frequency
of a cylindrical-shaped electron beam flowing inside of a cylindrical
tunnel, based on results obtained from Particle-in-cell (PIC) simulations.
In PIC simulations, we modulate the electron beam using two parallel,
non-intercepting, closely-spaced grids which are electrically connected
together by a single-tone sinusoidal voltage source. The electron
energy and the beam current distributions along the length of the
tunnel are monitored after the system is operating at steady-state.
We build a system matrix describing the beam's dynamics, estimated
by fitting a $2\times2$ matrix that best agrees with the first order
differential equations that govern the physics-based system. Results
are compared with the theoretical Branch and Mihran model, which is
typically used to compute the plasma frequency reduction factor in
such systems. Our method shows excellent agreement with the theoretical
model, however, it is also general. Our method can be potentially
utilized to determine the reduced plasma frequencies of electron beams
propagating in differently-shaped beam tunnels, where no theoretical
model yet exists, such as the case of a cylindrical or elliptical
electron beam propagating inside of a metallic beam tunnel of cylindrical,
square, or elliptical cross-section. It can be applied also to electron
beams composed of multiple streams.
\end{abstract}

\begin{IEEEkeywords}
Plasma frequency, particle-in-cell (PIC), electron beam, exceptional
points, degeneracy.

{\let\thefootnote\relax\footnotetext{This material is based upon work supported by the Air Force Office of Scientific Research award number FA9550-18-1-0355 and by the MURI Award number FA9550- 20-1-0409 administered through the University of New Mexico.}} 
\end{IEEEkeywords}

\vphantom{}

\section{Introduction}

The plasma frequency concept originates from the fact that an infinite
cloud of electrons with volumetric charge density $\rho_{v}$ oscillates
at a plasma frequency $\omega_{p}=\sqrt{\eta\rho_{v}/\varepsilon_{0}}$
when any electron in the cloud is perturbed from its equilibrium position,
where $\eta$ is the charge-to-mass ratio of an electron and $\varepsilon_{0}$
is the permittivity of free space \cite{tonks1929oscillations}. Plasma
oscillations in linear electron beams are induced by space-charge
fields, as explained in \cite{tsimring2006electron_ch7} and Ch. 9
of \cite{gilmour2011klystrons}, which result in repulsive longitudinal
forces between charges. However, the calculation of the plasma frequency
for a linear stream of charges with finite cross-section requires
the consideration of the radial variation of the space-charge fields,
electron velocities, and volume charge densities within the beam cross
section. Such parameters will fringe, or decay, with radial distance
away from the beam center due to the boundary between a beam of finite
cross-section and surrounding vacuum, as well as the presence of the
beam tunnel's conducting walls where the longitudinal electric field
vanishes. The electronic wave theory for linear beam tubes was developed
by Hahn \cite{hahn1939small} and refined by Ramo \cite{ramo1939space},
utilizing the solutions of Maxwell\textquoteright s equations with
corresponding boundary conditions to directly compute the reduced
plasma frequency of an electron beam that has a finite cross section
and is contained within a cylindrical metallic tunnel. Branch and
Mihran further built upon Ramo's work and introduced the plasma frequency
reduction factor term, $R$ for cases of both solid and annular electron
beams propagating within cylindrical metallic tunnels \cite{branch1955plasma}.
It was found that an electron beam with finite cross section will
have a plasma frequency that is reduced compared to that of an electron
beam with infinite cross section and with the same volumetric charge
density. Thus, the reduced plasma frequency is expressed as $\omega_{q}=R\omega_{p}$,
where $0<R<1$. The reduced plasma frequency is one of the fundamental
parameters that affects the dispersion characteristics of the so-called
space-charge wave or electronic waves, as explained in \cite{ramo1939space},
\cite{tsimring2006electron_ch7}, Ch. 9 in \cite{gilmour2011klystrons},
and Ch.9 in \cite{gewartowski1965principles}, which are the waves
within the electron beam that possess both velocity and charge modulation
in space and time. A linear stream of electrons with average speed
$u_{0}$ supports multiple space-charge waves with wave function $e^{j\omega t-jkz}$.
Typically, there are two dominant charge waves with approximate wavenumbers
$k=(\omega\pm\omega_{q})/u_{0}$ \cite{ramo1939space}, \cite{tsimring2006electron_ch7}.
Further theoretical effort was done in works \cite{datta2009simple,antonsen1998traveling},
where the authors came up with closed-form formulas to find the plasma
frequency reduction factor without numerically solving Branch and
Mihran's transcendental equation for the parameter $T$ within the
electron beam.

The plasma frequency reduction factor $R$ is one of the fundamental
design parameter in electron beam devices since it affects synchronization.
As a consequence, the value of $R$ affects the frequency of peak
gain in traveling wave tube amplifiers modeled by Pierce theory. Additionally,
the choice of drift-tube length between cavities in klystrons for
optimal extraction of energy from the modulated electron beam depends
on accurate computation of the reduced plasma frequency \cite{ramo1939space}.
The accurate calculation of $R$ is necessary for the modeling and
designing of TWTs . For instance, the famous classical theory developed
by Pierce requires the calculation of the reduced plasma frequency
in order to calculate the parameter $4QC^{3}=\omega_{q}^{2}/\omega^{2}$
\cite{pierce1950traveling}, Ch. 10 in \cite{gewartowski1965principles},
Ch. 12 in \cite{gilmour2011klystrons}, which is necessary to accurately
calculate the gain. However, the theoretical calculation of the reduced
plasma frequency loses accuracy when the tunnel geometry is different
from a circular cylinder or when the electron beam has a cross section
that is not circular or annular. The reduced plasma frequency for
beams in complex tunnel structures such as a cylindrical beam in folded
or helical waveguide is approximated in many works by assuming that
the tunnel is cylindrical, i.e., by neglecting the structure periodicity
and deformations in the shape of the cylindrical tunnel due to the
slow wave structure geometry. Some authors have previously provided
a formulation to determine the reduced plasma frequency of an electron
beam propagating within a helix slow-wave structure approximated using
the sheath helix model, as in \cite{antonsen1998traveling}; this
formulation has been used in TWT modeling software such as CHRISTINE
and LMsuite \cite{antonsen1997christine,wohlbier2002multifrequency}.
However, it is not trivial to compute $R$ using the sheath helix
model, and most authors studying linear beam tubes simply use the
values for $R$ from the Branch and Mihran model \cite{tien1955large,simon2017evaluation,wong2020recent}.
Furthermore, experimental work has also been performed in \cite{branch1967space,vorob1990experimental}
aiming at determining the reduced plasma frequency in linear beam
tubes. To date, there is a lack of literature available to provide
a robust method that can be easily used to verify or find the reduced
plasma frequency in complex structures.

In this paper, we present a method that is more general than the Branch
and Mihran model (which only yields real-valued space charge wavenumbers)
and is rather simple to use. We model the electron beam dynamics by
finding the system matrix that describes the small-signal evolution
of the electron beam with position and time. The method is the similar
to the the one presented in \cite{mealy2020traveling} that was applied
to traveling wave tubes. Here, the method is applied to find the eigenmodes
of the space-charge waves in a beam tunnel, i.e., waves that do not
interact with a propagating electromagnetic mode in a slow wave waveguide.
The presented method is based on defining and then finding the system
matrix through the interpretation of data extracted from particle-in-cell
(PIC) simulations. After the system matrix is determined, we then
calculated the two charge-wave wavenumbers. The reduced plasma frequency
of the electron beam is inferred by finding the deviation of the two
space charge wavenumbers from the average electronic phase constant
$\beta_{0}=\omega/u_{0}$.

The purpose of this paper is to show how the proposed method based
on three-dimensional PIC simulations is used to calculate the complex
wavenumbers of the space-charge waves supported by an electron beam.
Since this is the first time we apply the proposed method to the study
of the reduced plasma frequency, we apply it to the case of a circular
cylindrical beam within a circular cylindrical tunnel, i.e., the case
where the solution is known from a previously developed analytical
method \cite{ramo1939space,branch1955plasma,datta2009simple}. In
other words, the scope of this paper is to show that the proposed
method works for a simple configuration; hence it is a good candidate
to study even more complicated configurations where there is no known
analytical solution for the plasma frequency reduction factor, such
as the case of a two-stream electron beam propagating within a metallic
structure that can exhibit growing space charge waves (complex space
charge wavenumbers) due to the two-stream instability effect \cite{pierce1949dotlble,pierce1949new,figotin2021analytic_ch48}.
Motivated by previous work on two-stream instability amplifiers, Islam
et al., \cite{islam2022modeling,islam2022multiple} have proposed
a method to generate a two-stream electron beam with two different
energies from a single-cathode potential. Our PIC-based model may
be useful also to determine the complex space charge wavenumbers in
such a configuration. Additionally, our PIC-based model may also be
used to study the space-charge wavenumbers for electron beams which
interact with lossy materials, such as the case of the resistive wall
amplifier \cite{birdsall1953resistive,rowe2016metamaterial,rowe2015metamaterial,forbes2021effective}.

\begin{figure}
\begin{centering}
\centering \subfigure[]{\includegraphics[width=1\columnwidth]{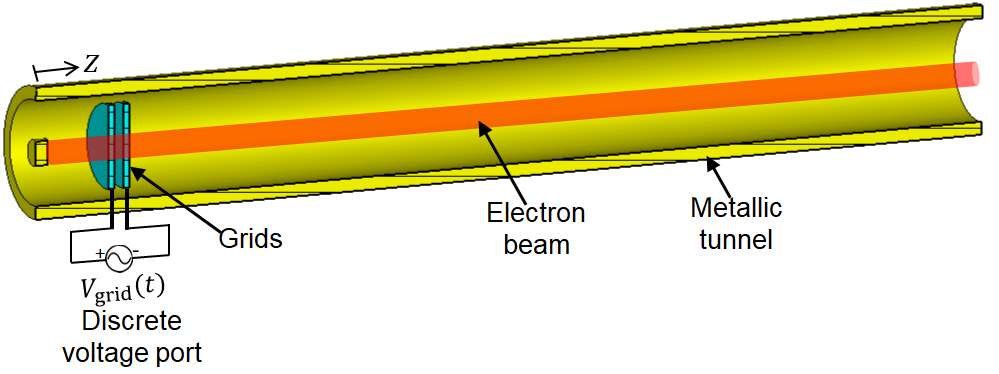}\label{Fig:General_Setup_a}}
\par\end{centering}
\begin{centering}
\centering \subfigure[]{\includegraphics[width=1\columnwidth]{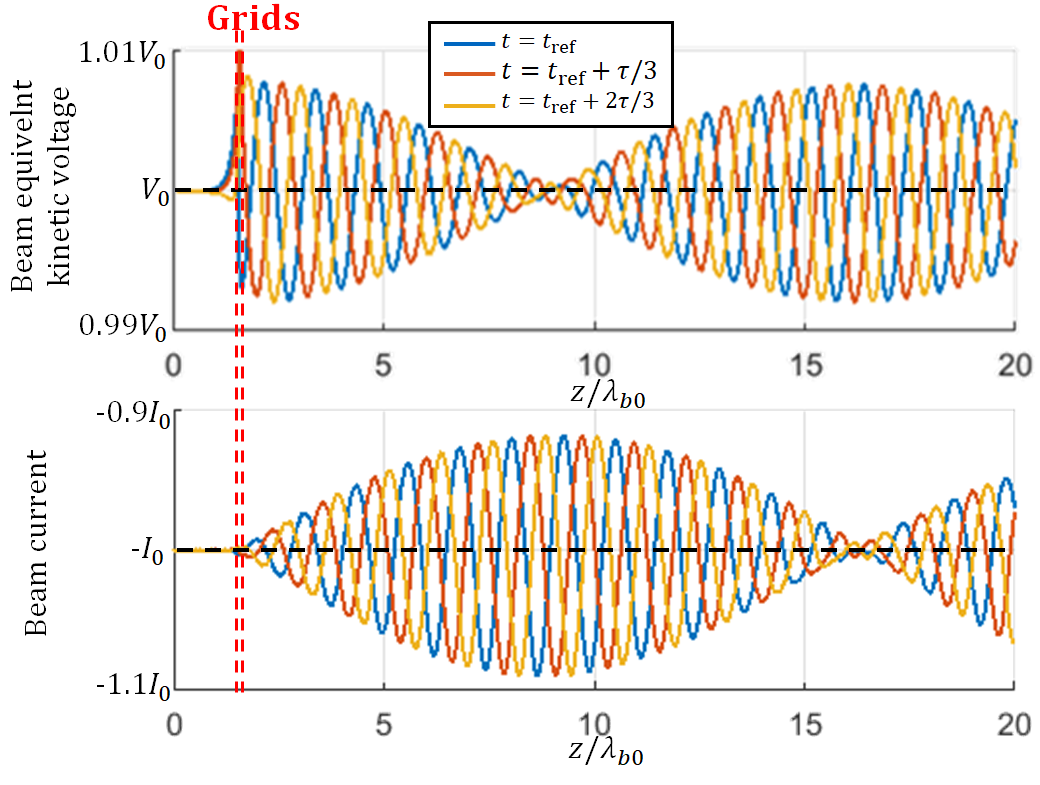}\label{Fig:Sample_Voltage}}
\par\end{centering}
\centering{}\caption{(a) Setup for PIC simulation used to determine the reduced plasma
frequencies of the electron beam in a cylindrical tunnel. Modulating
grids are simulated as perfect electric conductors which are transparent
to particles. (b) An example of the distribution of the electron beam
total (dc and ac) equivalent kinetic voltage, defined as $v_{b}^{tot}(z,t)=V_{0}+v_{b}(z,t)$
with dc equivalent kinetic voltage $V_{0}$ and ac equivalent kinetic
voltage $v_{b}(z,t)$ under the small-signal approximation, and current
$i_{b}^{tot}(z,t)=-I_{0}+i_{b}(z,t)$, calculated based on particles'
data exported from PIC simulations at steady state, showing the space-time
modulation in the beam voltage and current with respect to arbitrary
reference time $t_{\mathrm{ref}}$, where $\tau$ is the period of
the sinusoidal excitation and $\lambda_{b0}$ is the average electronic
wavelength. The knowledge of the electron beam voltage and current
is then used to estimate the $2\times2$ system matrix $\mathbf{M}$
that describes the differential equation governing the beam dynamics.
Finally, the system matrix is used to find the wavenumbers of space-charge
waves and consequently the reduced plasma frequency.}
\end{figure}

\section{PIC-based method}

The electron beam has equivalent kinetic dc voltage and dc current
$V_{0}$ and $I_{0}$, respectively, where the dc equivalent kinetic
voltage in the non relativistic case is $V_{0}=u_{0}^{2}/(2\eta)$
and $u_{0}$ is the electrons time-averaged speed, $c$ is the free
space speed of light, and $\eta=e/m$ is the charge to mass ratio
and $e$ is the electron charge represented as positive number. The
magnetically-confined electron beam has a circular cross-section with
radius $R_{b}$, and is flowing within a metallic cylindrical tunnel
with radius $R_{t}$, as illustrated in Fig. \ref{Fig:General_Setup_a}.
We introduce modulation to the beam using two closely-spaced grids
that are made of fictitious metal that is transparent to electrons,
i.e., it allows electrons to pass through the grid without being intercepted,
while preserving other properties a perfect electrical conductor.
The gap between the two grids is chosen to be very small compared
to wavelength $d_{grid}\ll\lambda_{0}$ to have a uniform electric
field distribution along the gap and to avoid the transit-time effects
for electrons. We apply an ac voltage $V_{grid}$ with a monochromatic
sinusoidal signal between the grids in order to generate an axial
electric field that modulates the beam in a simple and reproducible
way. We rely on three-dimensional PIC simulations implemented in CST
Particle Studio to satisfy the equations for charged particles motion
and Maxwell equations, which are discretized in space and time in
the PIC software.

The PIC solver calculates the instantaneous speed $u_{b}^{tot}(z,t)=u_{0}+u_{b}(z,t)$,
and location of discrete charged macroparticles, where $u_{b}(z,t)$
represents the ac component of velocity. We represent the total (including
dc and ac parts) equivalent kinetic voltage of a non relativistic
electron beam using a one dimensional (1D) function as $v_{b}^{tot}(z,t)=\left[u_{b}^{tot}(z,t)\right]^{2}/(2\eta)$
which is expressed as $v_{b}^{tot}(z,t)=V_{0}+v_{b}(z,t)$, where
$v_{b}(z,t)\approx u_{0}u_{b}(z,t)/\eta$ represents the ac modulation
function. Analogously, the total beam current is $i_{b}^{tot}(z,t)=-I_{0}+i_{b}(z,t)$,
where$i_{b}(z,t)$ represents the ac modulation function, as was done
in \cite{mealy2020traveling}. Note that the method can be applied
also to relativistic beams when using $V_{0}=c^{2}/\eta\left(\sqrt{1-(u_{0}/c)^{2}}-1\right)$
and $v_{b}^{tot}(z,t)=c^{2}/\eta\left(\sqrt{1-(u_{b}^{tot}/c)^{2}}-1\right)$,
and the ac part is defined as $v_{b}(z,t)=v_{b}^{tot}(z,t)-V_{0}$
. However relativistic beams are not considered in this paper and
the accuracy of the presented method in those cases is left to future
studies.

We define a state vector that describes the ac electron beam velocity
and current dynamics as

\begin{equation}
\mathbf{\boldsymbol{\psi}}(z,t)=[\begin{array}{cc}
v_{b}(z,t), & i_{b}(z,t)\end{array}]^{T}.\label{eq:TD-StateVect}
\end{equation}
We show in Fig. \ref{Fig:Sample_Voltage} an example of the distribution
of the electron beam total (ac and dc) equivalent kinetic voltage
and current calculated based on the particle's data exported from
PIC simulations at steady state showing the modulation in the beam
voltage and current due to the grids, where $\lambda_{0b}=u_{0}/f$.

At steady state, in the small-signal approximation, the state vector
is monochromatic with angular frequency $\omega$ and therefore it
is represented in phasor domain as \cite{mealy2020traveling}

\begin{equation}
\mathbf{\Psi}(z)=[\begin{array}{cc}
V_{b}(z), & I_{b}(z)\end{array}]^{T}.\label{eq:FD-StateVect}
\end{equation}
Because the problem we consider is uniform in the $z$-direction (due
to confinement of the beam by a strong axial magnetic field), we assume
that the evolution of the electron beam dynamic along the $z$-direction
is described by a first order differential equation as

\begin{equation}
\dfrac{d\mathbf{\Psi}(z)}{dz}=-j\mathbf{M}\mathbf{\Psi}(z),\label{eq:DE}
\end{equation}
where $\mathbf{M}$ is the $2\times2$ system matrix and the $e^{j\omega t}$
time dependence is assumed.

The state vector is calculated from the data exported from PIC simulation
as done in \cite{mealy2020traveling}. We calculate the state vector
at discrete locations $\Psi(z=n\Delta_{s})=\Psi_{n}$, where $n=0,1,..N$,
$\Delta_{s}$ is chosen to be a small position step size ($\Delta_{s}\leq\lambda_{0b}/3$
and $\lambda_{0b}=u_{0}/f$ ), and $N\Delta_{s}$ is the total length
of the structure. According to (\ref{eq:DE}), the sampled state vector
should satisfy the relations

\begin{equation}
\begin{array}{c}
\mathbf{\Psi}_{2}=\mathbf{T}\mathbf{\Psi}_{1},\ \ \ \ \ \ (\ref{eq:Tmatrix-StateVector}.1)\\
\mathbf{\Psi}_{3}=\mathbf{T}\mathbf{\Psi}_{2},\ \ \ \ \ \ (\ref{eq:Tmatrix-StateVector}.2)\\
\vdots\ \ \ \ \ \ \ \ \ \ \ \ \ \ \ \\
\mathbf{\Psi}_{N+1}=\mathbf{T}\mathbf{\Psi}_{N},\ \ \ \ \ \ (\ref{eq:Tmatrix-StateVector}.N)
\end{array}\label{eq:Tmatrix-StateVector}
\end{equation}
where $\mathbf{T}=e^{-j\mathbf{M}\Delta_{s}}$ is the transfer matrix,
$\mathbf{M}$ is related to $\mathbf{T}$ as $\mathbf{M}=j\ln(\mathbf{T})/\Delta_{s}$,
and the state vectors $\Psi_{n}$ are calculated directly from PIC
simulations.

The relations in (\ref{eq:Tmatrix-StateVector}) represent $2N$ linear
equations in $4$ unknowns, which are the elements of the transfer
matrix $\mathbf{T}$. Assuming $N>2$, the system in (\ref{eq:Tmatrix-StateVector})
is mathematically referred to as overdetermined because the number
of linear equations ($2N$ equations) is greater than the number of
unknowns (4 unknowns). An approximate solution that best satisfies
all the given equations in Eq. (\ref{eq:Tmatrix-StateVector}), i.e.,
minimizes the sums of the squared residuals, $\sum_{n}\left|\left|\mathbf{\Psi}_{n+1}-\mathbf{T}\mathbf{\Psi}_{n}\right|\right|^{2}$
is determined like in \cite{williams1990overdetermined,mealy2020traveling,anton2013elementary_ch4}
and is given by

\begin{equation}
\mathbf{T}_{\mathit{\mathrm{best,approx.}}}=\left(\mathbf{W}_{2}\mathbf{W}_{1}^{T}\right)\left(\mathbf{W}_{1}\mathbf{W}_{1}^{T}\right)^{-1},\label{eq:Best_Fit}
\end{equation}
 where
\begin{equation}
\mathbf{W}_{1}=\left[\begin{array}{cccc}
\mathbf{\Psi}_{1}, & \mathbf{\Psi}_{2}, & \ldots & \mathbf{\Psi}_{N}\end{array}\right],
\end{equation}
and

\begin{equation}
\mathbf{W}_{2}=\left[\begin{array}{cccc}
\mathbf{\Psi}_{2}, & \mathbf{\Psi}_{3}, & \ldots & \mathbf{\Psi}_{N+1}\end{array}\right],
\end{equation}
are $2\times N$ matrices.

Assuming the state vectors take the form of a wave function $\mathbf{\Psi}(z)\propto e^{-jkz}$,
\ref{eq:DE} is simplified to as $k\Psi=\mathbf{M}\mathbf{\Psi},$which
constitutes an eigenvalue problem. Therefore, the space-charge waves'
wavenumbers are the eigenvalues of $\mathbf{M}_{\mathrm{best}}$, 

\begin{equation}
k=\mathrm{eig}\left(\mathbf{M}_{\mathrm{best}}\right),\label{eq:eig_Mbest}
\end{equation}
where $\mathbf{M}_{\mathrm{best}}=j\ln(\mathbf{T}_{\mathit{\mathrm{best}}})/\Delta_{s}$
, which leads to two solutions: $k_{1}$ and $k_{2}$. The PIC-based
reduced plasma frequency is calculated as $\omega_{q,\mathrm{PIC}}=u_{0}\mathrm{Re}(k_{2}-k_{1})/2$.
The associated ``PIC-based'' reduction factor calculated as

\begin{equation}
R=\omega_{q\mathrm{,PIC}}/\omega_{p}\label{eq:ReductionFactPIC}
\end{equation}
where $\omega_{p}=\sqrt{\eta I_{o}/(Au_{0}\varepsilon_{0})}$ and
the beam has cross-sectional area $A=\pi R_{b}^{2}$.

\section{Illustrative example }

As an illustrative example, we consider a solid electron beam with
equivalent kinetic dc voltage $V_{0}=6.7\ \mathrm{kV}$ (which corresponds
to $u_{0}=\sqrt{2\eta V_{0}}=0.16c$), beam radius $R_{b}=0.5\ \mathrm{mm}$,
and the beam current $I_{0}$ that is swept. The tunnel is made of
a perfect electric conductor and has radius $R_{t}=2\ \mathrm{mm}$
and total length of $120\ \mathrm{mm}$. We use two grids spaced apart
with a gap of $d_{\mathrm{grid}}=0.2\ \mathrm{mm}$ and a grid excitation
voltage $V_{\mathrm{grid}}(t)=100\cos\left(2\pi ft\right)\ \mathrm{volts}$.
An axial dc magnetic field of 1 T is used to confine the electron
beam. All 3D PIC simulations in this paper are performed using CST
Particle Studio. The 3D segmentation performed in CST uses hexahedral
mesh with mesh size of approximately $\Delta_{\mathrm{mesh}}$=$\lambda_{b0}/30=0.2$
mm calculated at $f=10$ GHz, where $\lambda_{b0}=u_{0}/f$ and $u_{0}=0.16c$.
The total number of charged particles used by PIC simulations to model
the electron beam is approximately $10^{6}$ particles. We run PIC
simulations for a total time of $t_{\mathrm{sim}}=$10 ns, where the
beam dynamics in phasor domain are obtained based on the particles'
data on the time period from $t=t_{\mathrm{sim}}-\tau\to t_{\mathrm{sim}}$,
where $\tau=1/f$ is the period of the applied sinusoidal signal.

\begin{figure}
\begin{centering}
\centering \subfigure[]{\includegraphics[width=1\columnwidth]{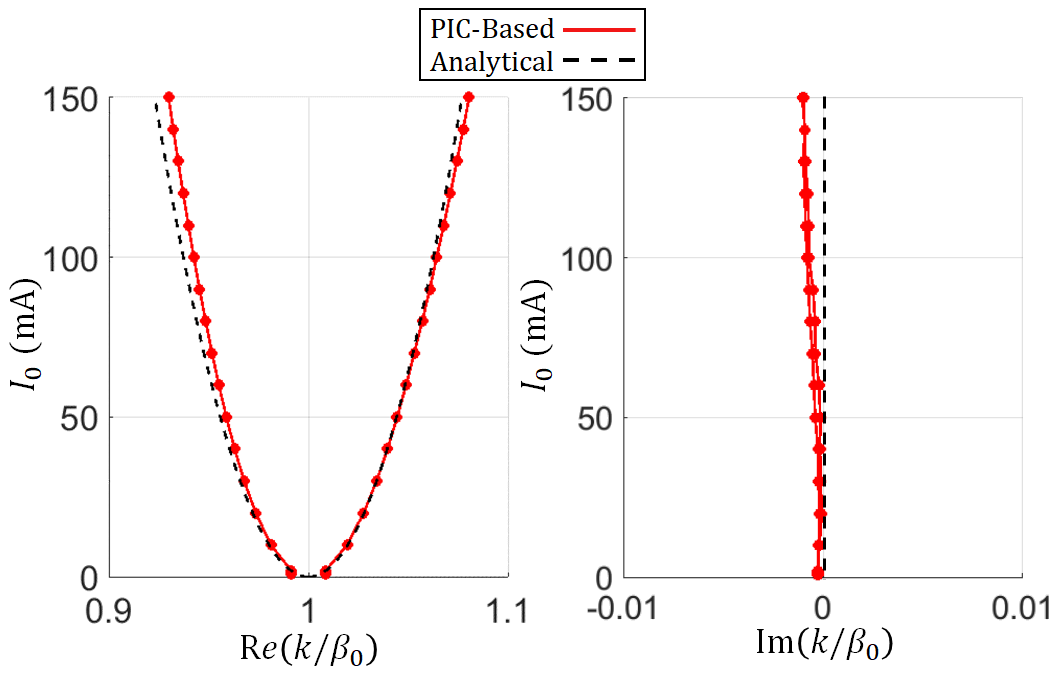}\label{Fig:Disp_Sweep_I0}}
\par\end{centering}
\begin{centering}
\centering \subfigure[]{ \includegraphics[width=1\columnwidth]{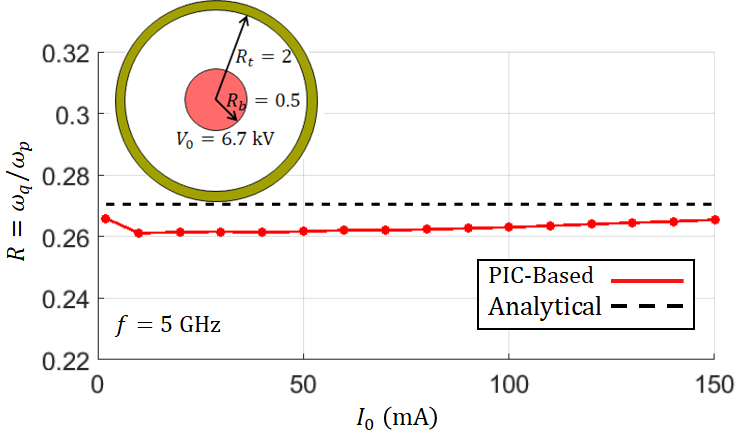}\label{Fig:R_Sweep_I0}}
\par\end{centering}
\centering{}\caption{(a) Dispersion relation showing the wavenumbers of the eigenmodes
of the space-charge wave supported by the electron beam. Dimensions
are shown in the inset of (b), with units of mm. The proposed PIC-based
method is compared with the analytical results from Branch and Mihran
\cite{branch1955plasma}. (b) Corresponding plasma frequency reduction
factor versus beam current. Our results are in agreement with the
analytical ones in (\ref{eq:R_theory}) and (\ref{eq:ramo_Disp}),
that show that the plasma frequency reduction factor is current independent. }
\end{figure}

We sweep the electron beam dc current $I_{0}$ at constant frequency
of $f=5$ GHz. We show in Fig. \ref{Fig:Disp_Sweep_I0} the ``PIC-based''
dispersion relation for the space-charge wave showing the complex-valued
wavenumbers $k_{1}$ and $k_{2}$ for the two space-charge modes versus
beam current, calculated as the eigenvalues for the system matrix
obtained based on PIC simulations, $\mathbf{M}_{\mathrm{best}}$.
Since we have lossless metals and vacuum in this example, the small
imaginary part ($\left|\mathrm{Im}(k)/\mathrm{Re}(k)\right|<10^{-3}$)
of the wavenumbers shown in Fig. \ref{Fig:Disp_Sweep_I0} may be attributed
to numerical error in our PIC-based method. The method used to find
the theoretical wavenumber (dashed-black curves in Fig. \ref{Fig:Disp_Sweep_I0})
is based on the work by Branch and Mihran \cite{branch1955plasma},
which gives purely real wavenumbers. The dispersion diagram in Fig.
\ref{Fig:Disp_Sweep_I0} shows a good match between the real value
of the space-charge wavenumber calculated based on the proposed method
(based on Eq. \ref{eq:eig_Mbest}) and the one calulated based on
Branch and Mihran's method \cite{branch1955plasma} (which is real-valued).
We show in Fig. \ref{Fig:R_Sweep_I0} the ``PIC-based'' reduction
factor calculated as in (\ref{eq:ReductionFactPIC}).

\begin{figure}
\begin{centering}
\centering \subfigure[]{\includegraphics[width=1\columnwidth]{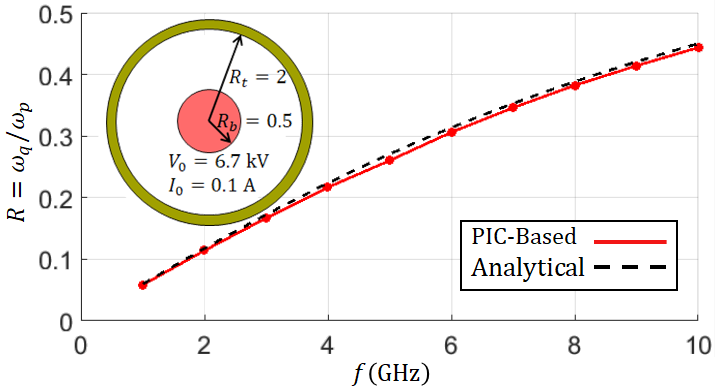}\label{Fig:R_Sweep_freq}}
\par\end{centering}
\begin{centering}
\centering \subfigure[]{ \includegraphics[width=1\columnwidth]{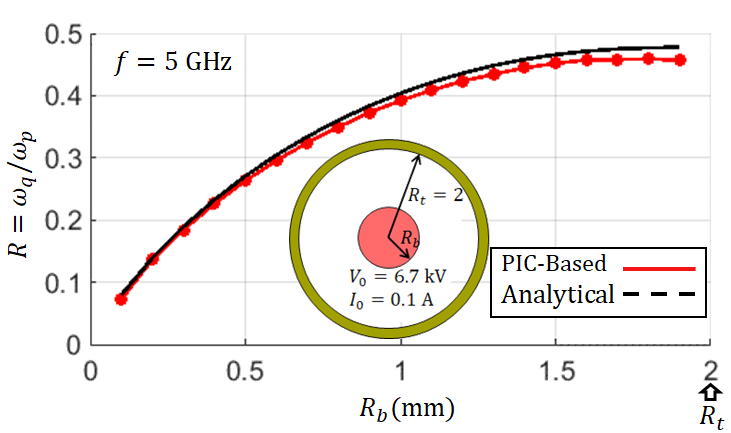}\label{Fig:R_Sweep_Rb}}
\par\end{centering}
\centering{}\caption{Reduction factor calculated for a cylindrical beam in a metallic tunnel
with geometry as shown in the insets, when (a) the frequency is swept,
and (b) the beam radius is swept. The dimensions shown in the insets
are in mm.}
\end{figure}

We now compare the PIC-based result obtained from the method described
above with the analytical one based on Branch and Mihran's work in
\cite{branch1955plasma}. According to that theory, the plasma frequency
reduction factor is approximated as

\begin{equation}
\begin{array}{c}
R_{\mathrm{Theory}}=\dfrac{1}{\sqrt{1+\left(T/\beta_{0}\right)^{2}}},\end{array}\label{eq:R_theory}
\end{equation}
where $\beta_{0}=\omega/u_{0}$ is the mean electronic phase constant
and the parameter $T$ is found by solving the following nonlinear
equation \cite{ramo1939space,branch1955plasma}

\begin{equation}
TR_{b}\dfrac{J_{1}(TR_{b})}{J_{0}(TR_{b})}=\beta_{0}\dfrac{K_{0}(\beta_{0}R_{t})I_{1}(\beta_{0}R_{b})+K_{1}(\beta_{0}R_{b})I_{0}(\beta_{0}R_{t})}{K_{0}(\beta_{0}R_{b})I_{0}(\beta_{0}R_{t})-K_{0}(\beta_{0}R_{t})I_{0}(\beta_{0}R_{b})},\label{eq:ramo_Disp}
\end{equation}

\noindent where the various orders and kinds of Bessel functions are
defined in \cite{branch1955plasma}. The two real-valued wavenumbers
$k_{1}=\beta_{0}-\omega_{q}/u_{0}$ and $k_{2}=\beta_{0}+\omega_{q}/u_{0}$
based on Branch and Mihran's method \cite{branch1955plasma}, where
$\omega_{q}=R_{\mathrm{Theory}}\omega_{p}$, are plotted in Fig. \ref{Fig:Disp_Sweep_I0},
as ``Analytical'', and the reduction factor $R_{\mathrm{Theory}}$
is shown in Fig. \ref{Fig:R_Sweep_I0}. Note that from (\ref{eq:R_theory})
and (\ref{eq:ramo_Disp}), the reduction factor is independent of
the beam current value $I_{0}$. However, the unreduced plasma frequency
and the two wavenumbers are all dependent on beam current. As expected,
the reduction factor obtained by the proposed PIC-based method seems
to be almost constant when varying the beam dc current. The results
show a good match between the plasma frequency reduction factor obtained
based on PIC simulations and that based on theoretical formulas in
(\ref{eq:R_theory}) and (\ref{eq:ramo_Disp}), with a maximum absolute
error in $R$ of 0.01, which is very small with respect to the reduction
factor obtained theoretically. Note that when $I_{0}\to0$, the plasma
frequency $\omega_{p}\to0$ and therefore the two space-charge waves
have wavenumbers $k_{1}=k_{2}=\beta_{0}$ as illustrated in Fig \ref{Fig:Disp_Sweep_I0}.
The dispersion relation around $I_{0}=0$ can be approximated to a
quadratic polynomial as $(k-\beta_{0})^{2}\propto I_{0}$ as illustrated
in Fig \ref{Fig:Disp_Sweep_I0}, and this has a deep physical meaning
as it will be discussed later on. We show in Fig. \ref{Fig:R_Sweep_freq}
and Fig. \ref{Fig:R_Sweep_Rb} the plasma frequency reduction factor
when the frequency is swept at constant beam dc current of $I_{0}=100$
mA, and when the beam radius is swept at constant dc current $I_{0}=100$
mA and frequency $f=5$ GHz, respectively. The agreement between the
PIC-based solution provided in this paper and the analytical one from
(\ref{eq:R_theory}) and (\ref{eq:ramo_Disp}) is excellent.

\section{Degeneracy of the spectrum }

To provide a physical insight into the result, it is convenient to
analyze the spectral properties of the system matrix $\mathbf{M}$
that describes the system as in (\ref{eq:DE}). Following the space-charge
wave theory in \cite{tsimring2006electron_ch7} and also using the
formulation presented in \cite{rouhi2021exceptional}, we use the
analytically determined matrix

\noindent 
\begin{equation}
\mathbf{M}_{\mathrm{Theory}}=\left[\begin{array}{cc}
\beta_{0} & \dfrac{R_{\mathrm{Theory}}^{2}}{\left(A\omega\varepsilon_{0}\right)}\\
\beta_{p}^{2}\left(A\omega\varepsilon_{0}\right) & \beta_{0}
\end{array}\right],\label{eq:matrix_Charge}
\end{equation}

\noindent where $\beta_{p}=\omega_{p}/u_{0}$ is the unreduced plasma
phase constant, and the reduction factor $R_{\mathrm{Theory}}$ is
taken from (\ref{eq:R_theory}). The theoretical wavenumbers of the
space-charge waves are found as the eignvalues of the system matrix
$\mathbf{M}_{\mathrm{Theory}}$, $k_{1}=\beta_{0}-\omega_{q}/u_{0}$
and $k_{2}=\beta_{0}+\omega_{q}/u_{0}$ which agree with the space-charge
wavenumber expressions found in \cite{ramo1939space,branch1955plasma,datta2009simple},
and correspond to those already plotted in Fig. \ref{Fig:Disp_Sweep_I0}.
The ac electron kinetic energy and the ac beam current of the two
space-charge waves are described by the two eigenvetors of the matrix
(\ref{eq:matrix_Charge}) as $\Psi_{1}=\left[\begin{array}{cc}
-R_{\mathrm{Theory}}/\left(\beta_{p}A\omega\varepsilon_{0}\right), & 1\end{array}\right]^{T}$ and $\Psi_{2}=\left[\begin{array}{cc}
R_{\mathrm{Theory}}/\left(\beta_{p}A\omega\varepsilon_{0}\right), & 1\end{array}\right]^{T}$.

\begin{figure}
\begin{centering}
\includegraphics[width=1\columnwidth]{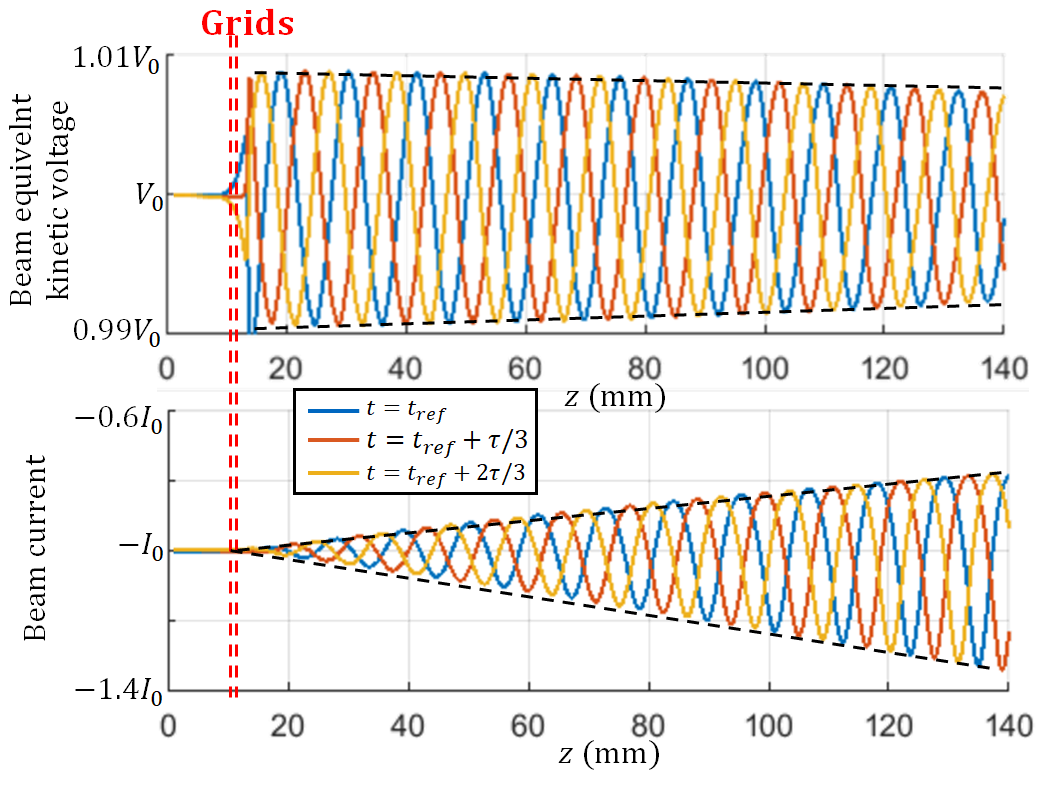}
\par\end{centering}
\begin{centering}
\label{Fig:EPD_Charge_Wave}
\par\end{centering}
\centering{}\caption{An example of electron beam dynamics when the system is very close
to an EPD. The distribution of the electron beam total equivalent
kinetic voltage and current decays and grows algebraically, respectively,
along the $z$ direction. }
\end{figure}

We know that when two eigenvectors coalesce, the system experiences
an exceptional point of degeneracy (EPD) of order 2, where the system
matrix is not diagonalizable and is instead similar to a Jordan block
of order 2, as explained in \cite{mealy2019exceptional,rouhi2021exceptional,Figotin21ExceptionalP,kato2013perturbation_ch2}.
Therefore, the eigenmodes experience an algebraic linear behavior,
besides the usual phase propagation along $z$. At an EPD where the
space charge waves are degenerate (i.e., when $k_{1}=k_{2}=\beta_{0}$),
the beam ac voltage is expressed as $V_{b}(z)=(u_{1}+u_{2}z)e^{-j\beta_{0}z}$,
whereas away from the EPD, it is expressed as $V_{b}(z)=e^{-j\beta_{0}z}(u_{1}e^{-j\beta_{q}z}+u_{2}e^{j\beta_{q}z})$.
The point $I_{0}=0$ in Fig. \ref{Fig:Disp_Sweep_I0} represents an
EPD, and indeed, in its proximity the two wavenumbers follow the law
$(k-\beta_{0})^{2}\propto I_{0}$. A vanishing $I_{0}$ indicates
the absence of the beam but one can still see the physics pertaining
to an EPD at regimes where $I_{0}$ is very small. The two eigenvectors
also coalesce when the reduction factor $R$ is very small, hence
the system can experiences and EPD at $R=0$. It is convenient to
use a ``coalescence parameter'' that quantifies the vicinity of
the two eigenvector to each other (i.e., to describe how close is
a system's regime to an EPD). To check the coalescence parameter,
we look at the angle between the two system's two eigenvectors, when
the current elements are scaled by an impedance of $Z_{0}=\beta_{0}/\left(A\omega\varepsilon_{0}\right)$
to have vectors that have all elements in volts, that is defined by
the normalized scalar product as $\cos\theta_{12}=\mathrm{Re}\left(\Psi_{1}\cdot\Psi_{2}\right)/\left(\parallel\Psi_{1}\parallel\,\parallel\Psi_{1}\parallel\right)$.
The coalescence parameter is here calculated as

\begin{equation}
\sin(\theta_{12})=\dfrac{2R\left(\dfrac{\omega_{p}}{\omega}\right)}{1+R^{2}\left(\dfrac{\omega_{p}}{\omega}\right)},
\end{equation}
 and it describes how close the two eigenvectors of the system are
to coalescing (i.e., when the systems experiences an EPD), which occurs
when $\sin\theta_{12}=0$, i.e., when either $R=0$ or $\omega_{p}=0$.
Indeed, in regimes where the plasma frequency is very small with respect
to the operating frequency, the system is very close to an EPD. We
show in Fig. \ref{Fig:EPD_Charge_Wave} the space-charge wave behavior
when the system is very close to EPD, i.e., for a case where the plasma
frequency is $f_{p}=0.11$ GHz and the operating frequency $f=$5
GHz, where a beam dc current $I_{0}=1$ mA is used. The ac beam equivalent
voltage and current decays and grows, respectively, linearly, along
$z$. This behavior is different from other cases where the EPD does
not occur, as shown in Fig. \ref{Fig:Sample_Voltage} for example.

\section{Conclusion}

A method to determine the reduction factor for single stream electron
beam flowing inside of a tunnel has been demonstrated using a novel
technique. Our method is based on analyzing the data obtained from
time-domain 3D PIC simulations, hence it accounts for all the physical
aspects of the problem. Our model is general and can be applied to
several other geometries supporting electron beams that are different
from the one used here for demonstration purposes. The proposed method
seems precise since the calculated reduction factor is in good agreement
with the one obtained analytically for cylindrically-shaped electron
beam flowing inside a cylindrical metallic tunnel \cite{branch1955plasma}.
Since the proposed method is just based on 3D PIC simulations, it
can be utilized to study electron beams in complex-shaped beam tunnels,
where no theoretical model yet exists, or even for electron beams
made of multiple streams as in \cite{islam2022modeling,islam2022multiple,figotin2021analytic_ch48}.

\section*{Acknowledgment}

The authors are thankful to DS SIMULIA for providing CST Studio Suite
that was instrumental in this study.

\bibliographystyle{ieeetr}
\bibliography{myref}

\end{document}